\documentclass{article}

\usepackage{amsmath,amssymb,graphicx,calc,capt-of,ifthen}

\newcommand{\asterisk}{*}
\newcommand{\tmop}[1]{\ensuremath{\operatorname{#1}}}

\newcommand{\tmtextit}[1]{{\itshape{#1}}}

\newenvironment{itemizedot}{\begin{itemize} }{\end{itemize}}
\newcommand{\tmfloatcontents}{}
\newlength{\tmfloatwidth}
\newcommand{\tmfloat}[5]{
  \renewcommand{\tmfloatcontents}{#4}
  \setlength{\tmfloatwidth}{\widthof{\tmfloatcontents}+1in}
  \ifthenelse{\equal{#2}{small}}
    {\ifthenelse{\lengthtest{\tmfloatwidth > \linewidth}}
      {\setlength{\tmfloatwidth}{\linewidth}}{}}
    {\setlength{\tmfloatwidth}{\linewidth}}
  \begin{minipage}[#1]{\tmfloatwidth}
    \begin{center}
      \tmfloatcontents
      \captionof{#3}{#5}
    \end{center}
  \end{minipage}}

\begin{document}

\title{Interacting Cosmological Fluids and the Coincidence Problem}
\author{Sean Z.W. Lip \thanks{S.Z.W.Lip@damtp.cam.ac.uk} \\
DAMTP, Centre for Mathematical Sciences,\\
Cambridge University, Wilberforce Road,\\
Cambridge CB3 0WA, UK $\ $}
\date{}
\maketitle

\begin{abstract}
  We examine the evolution of a universe comprising two interacting fluids,
  which interact via a term proportional to the product of their densities. In
  the case of two matter fluids it is shown that the ratio of the densities
  tends to a constant after an initial cooling-off period. We then obtain a
  complete solution for the cosmological constant $(w = - 1)$ scenario.
  Finally, we investigate the general case in which the dark energy equation
  of state is $p = w \rho$, where $w$ is a constant, and show that periodic
  solutions can occur if $w < - 1$. We further demonstrate that the ratio of
  the dark matter to dark energy densities is confined to a bounded interval, and
  that this ratio can be $O (1)$ at infinitely many times in the history of
  the universe, thus solving the coincidence problem.

  PACS: 98.80.-k, 95.36.+x
\end{abstract}

\begin{center}
\textbf{I. INTRODUCTION}
\end{center}

Observations based on Type Ia supernovae and the cosmic microwave background
suggest that the universe consists mainly of a non-gravitating type of matter 
called `dark energy', as well as a substantial amount of gravitating non-baryonic
`dark matter' [1--5]. Surprisingly, although the density of dark matter is expected to 
decrease at a faster rate than the density of dark energy throughout the history of
the universe, their magnitudes are comparable today. This is known as the `coincidence problem', 
and various attempts at its solution include the use of tracker fields [6] and oscillating 
dark energy models [7]. 

We will discuss a third possibility that has gained some attention 
recently, which is that dark energy and dark matter interact via an additional coupling 
term in the fluid equations [8--69]. The interaction is usually assumed to take the form
$A H \rho_m + B H \rho_{\Lambda}$, where $H$ is the Hubble parameter, $\rho_m$ and $\rho_{\Lambda}$
are the dark matter and dark energy densities respectively, and $A, B$ are dimensionless constants.
However, in this paper we will instead consider an interaction of the form 
$\gamma \rho_m \rho_{\Lambda}$, where $\gamma$ is a (non-dimensionless) constant [8]. Such an interaction 
is natural and physically viable, since we would expect the interaction rate to vanish if one of the densities 
is zero, and to increase with each of the densities. This form of interaction has also been used to
model systems ranging from standard two-body chemical reactions to predator-prey systems in biology.
Statistical fits to observed data suggest that this form of coupling helps
to alleviate the coincidence problem [9], and it has been shown that among holographic
dark energy models with an interaction term $\gamma \rho_m^{\alpha} \rho_{\Lambda}^{\beta}$ 
(where $\alpha, \beta \geqslant 0$ are integers), the one with $\alpha = \beta = 1$ gives the best 
fit to observations [10].

We propose to investigate this model in more detail. Our analysis will
consider the interaction of a dust fluid and a second fluid with an equation
of state $p = w \rho$, where $w$ is a constant. If the second fluid is also dust $(w = 0)$, we show
that the ratio of the densities tends to a constant after an initial
cooling-off period. Thus, matter fluids coupled in this way can be considered,
at late times, to evolve as a single non-self-interacting matter fluid. We
then go on to obtain a complete solution of the system for conventional $(p =
- \rho)$ dark energy, but show that such a model cannot address the
coincidence problem. Finally, we exhibit the various scenarios that can arise
for other forms of dark energy, and demonstrate that if $w < -1$ and there is an energy transfer
from dark energy to dark matter, periodic solutions can
arise in which the dark densities are comparable for a substantial fraction 
of the evolution of the universe. This would address the coincidence problem.

\begin{center}
\textbf{II. SETTING THE STAGE}
\end{center}

We work in a spatially flat Friedmann-Robertson-Walker universe, and assume
that it contains the following perfect fluids: dark matter, with density $\rho_m$, pressure $p_m$ and equation of state $p_m =
0$; `dark energy', with density $\rho_{\Lambda}$, pressure $p_{\Lambda}$ and 
an equation of state of the form $p_{\Lambda} = w \rho_{\Lambda} c^2$, where 
$w$ is a constant (in the case $w = -1$, we have a standard cosmological constant); 
baryonic matter, with density $\rho_b$, pressure $p_b$ and equation of state $p_b = 0$; 
and radiation, with density $\rho_r$, pressure $p_r$ and equation of state $p_r = \rho_r c^2 / 3$.

The conservation equations governing the evolution of these fluids are:
\begin{eqnarray}
  \dot{\rho_{}}_m & = & - 3 H \rho_m + \gamma \rho_m \rho_{\Lambda} \\
  \dot{\rho}_{\Lambda} & = & - 3 (1 + w) H \rho_{\Lambda} - \gamma \rho_m \rho_{\Lambda} \\
  \dot{\rho}_{b} & = & - 3 H \rho_b \\
  \dot{\rho}_{r} & = & - 4 H \rho_r \\
  3 H^2 & = & 8 \pi G (\rho_m + \rho_{\Lambda} + \rho_b + \rho_r)
\end{eqnarray}
where $H = \dot{a} / a$ is the Hubble parameter, and $a (t)$ is the
scale-factor. Here, an overdot indicates a derivative with respect to (cosmic)
time $t$. Note that the constant $\gamma$ is not dimensionless; it has dimensions 
of volume per unit mass per unit time. If $\gamma > 0$, energy is transferred from
dark energy to dark matter; the opposite occurs
if $\gamma < 0$. In the case $\gamma = 0$, the two fluids do not interact. In
[11], it is pointed out that $\gamma < 0$ would worsen the coincidence problem,
and in [12], it is argued that we need $\gamma > 0$ in order for the second
law of thermodynamics and Le Ch\^atelier's principle to hold. These results
are borne out later in Section 5 of this paper. Note that we do not include baryonic matter
in the interaction, due to the constraints imposed by local gravity measurements [70--71].

From the above equations, it is clear that $\rho_b$ and $\rho_r$ decrease as $a^{-3}$ and $a^{-4}$
respectively. In the next three sections of this paper, we will restrict ourselves to a late-time analysis,
where $\rho_b$ and $\rho_r$ are small and can be neglected. This leads to the reduced
set of equations:
\begin{eqnarray}
  \dot{\rho_{}}_m & = & - 3 H \rho_m + \gamma \rho_m \rho_{\Lambda} \\
  \dot{\rho}_{\Lambda} & = & - 3 (1 + w) H \rho_{\Lambda} - \gamma \rho_m \rho_{\Lambda} \\
  3 H^2 & = & 8 \pi G (\rho_m + \rho_{\Lambda})  \label{Hub}
\end{eqnarray}
For simplicity, we will also adopt units in which $8 \pi G = 1$ and $c = 1$ in these sections, which
contain a qualitative, mathematical analysis of the system. We will return to the 
full set of equations in Section VI when attempting to constrain the parameters 
of the system using observations.

Differentiating (\ref{Hub}) with respect to $t$ and substituting for
$\dot{\rho}_m$ and $\dot{\rho}_{\Lambda}$ gives the auxiliary equation
\begin{equation}
  2 \dot{H} = - \rho_m - (1 + w) \rho_{\Lambda} \label{Hub2}
\end{equation}
which will be useful later. Note that this equation is independent
of the coupling parameter $\gamma$; this is a consequence of energy
conservation.

It will be of interest to consider the acceleration of the universe, which is
given by the relation $\ddot{a} = a ( \dot{H} + H^2)$. From (\ref{Hub}) and
(\ref{Hub2}), we have
\begin{equation}
  \ddot{a} = - \frac{a}{6} (\rho_m + (1 + 3 w) \rho_{\Lambda}) . \label{ACCEL}
\end{equation}
Hence, since we assume that $a (t)$, $\rho_m$ and
$\rho_{\Lambda}$ are always non-negative, it is necessary (but not sufficient)
that $w < - \frac{1}{3}$ in order for the universe to accelerate today.

\begin{center}
\textbf{III. A SIMPLE CASE: TWO DUST FLUIDS}
\end{center}

We start by considering a model in which our fluids are both dust. The
equations of the system are:
\begin{eqnarray}
  \text{} \dot{\rho}_1 & = & - 3 H \rho_1 + \gamma \rho_1 \rho_2  \label{B1}\\
  \dot{\rho}_2 & = & - \gamma \rho_1 \rho_2 - 3 H \rho_2  \label{B2}\\
  3 H^2 & = & \rho_1 + \rho_2  \label{B3}
\end{eqnarray}
Define $u \equiv \rho_1 + \rho_2$. Then $\dot{u} = - \sqrt{3} u^{3 / 2}$, and
integrating gives $\frac{2}{\sqrt{u}} = \sqrt{3} (t + C)$, where $C$ is an
arbitrary constant. Therefore, $u = \frac{4}{3 (t + C)^2}$, and by shifting
the time coordinate ($t \rightarrow t - C$) we can set $C = 0$. Hence $H =
\frac{u}{\sqrt{3}} = \frac{2}{3 t}$ (as would be expected for a
matter-dominated universe).

Substituting this result into (\ref{B1}) and (\ref{B3}) yields the equation
\begin{eqnarray}
  \dot{\rho}_1 & = & - \frac{2}{t} \rho_1 + \gamma \rho_1 \left( \frac{4}{3
  t^2} - \rho_1 \right) 
\end{eqnarray}
which integrates to $\rho_1 = \frac{4}{t^2 (4 A e^{4 \gamma / 3 t} + 3)}$,
$\rho_2 = \frac{4}{t^2 (4 B e^{- 4 \gamma / 3 t} + 3)}$. The condition $\rho_1
+ \rho_2 = \frac{4}{3 t^2}$ leads to the requirement that $16 A B = 9$, which
leads to the general solution
\[ \rho_1 = \frac{4}{t^2 (4 A e^{4 \gamma / 3 t} + 3)}, \rho_2 = \frac{4}{t^2
   \left( \frac{9}{4 A} e^{- 4 \gamma / 3 t} + 3 \right)} . \]
The ratio of these densities is therefore
\[ \frac{\rho_1}{\rho_2} = \frac{3}{4 A} e^{- \frac{4 \gamma}{3 t}} \]
and, as $t \rightarrow \infty$, this tends exponentially quickly to the
constant $3 / 4 A$, with the individual densities evolving as $\rho_i \sim
t^{- 2}$.

This suggests that even if there were several types of `dust' in the universe
mutually interacting in this manner, we could approximate the evolution of
their densities by treating them as non-interacting after a short initial
period. (In other words, we might write $\rho_i \approx C_i t^{- 2}$ for $t >
t^{\asterisk}$, where the relative magnitudes of the $C_i$ are determined from
a relatively short initial evolution up to time $t^{\asterisk}$.)

\begin{center}
\textbf{IV. `CONVENTIONAL' DARK ENERGY ($w = -1$)}
\end{center}

We now consider a universe containing a dark matter fluid and a dark energy
fluid, in which the dark energy behaves like a cosmological constant (that is,
it satisfies the equation of state $p_{\Lambda} = - \rho_{\Lambda}$). The
dynamical equations simplify to:
\begin{eqnarray}
  \dot{\rho}_m & = & - 3 H \rho_m + \gamma \rho_m \rho_{\Lambda}  \label{A1}\\
  \dot{\rho}_{\Lambda} & = & - \gamma \rho_m \rho_{\Lambda}  \label{A2}\\
  H^2 & = & \frac{1}{3} (\rho_m + \rho_{\Lambda})  \label{A3}\\
  2 \dot{H} & = & - \rho_m  \label{A4}
\end{eqnarray}
In the non-interacting case $\gamma = 0$, it is easy to see that the dark
energy density $\rho_{\Lambda}$ stays constant, and $\dot{\rho}_m < 0$ (in
fact, $\rho_m \propto a^{- 3} \propto t^{- 2}$), so the dark matter becomes
more diffuse. The ratio $\frac{\rho_m}{\rho_{\Lambda}}$ decreases as $t^{-
2}$, and so the coincidence problem remains. Since $\rho_m \rightarrow 0$, we
can see from (\ref{ACCEL}) that the universe accelerates at an increasing rate
as time progresses.

Now suppose $\gamma \neq 0$. First, observe that $H = 0$ implies $\rho_m =
\rho_{\Lambda} = 0$, so we restrict attention to positive $H$. This implies an
eternally expanding universe. From (\ref{A2}) and (\ref{A4}) it follows that
$\rho_{\Lambda} = A e^{2 \gamma H}$, where $A$ is a positive constant. Given
$\gamma$, we can find $A$ based on the values of $\rho_{\Lambda}$ and $H$
observed today. Using (\ref{A3}), we can then parametrize the whole system in
terms of $H$:
\begin{eqnarray}
  \rho_m & = & 3 H^2 - A e^{2 \gamma H} \\
  \rho_{\Lambda} & = & A e^{2 \gamma H}  \label{SOL2}
\end{eqnarray}
From (\ref{SOL2}) we can see that $\rho_{\Lambda}$ never vanishes, so dark
energy is a perpetual component of the universe. Now, using (\ref{A4}), we can
obtain a first-order separable differential equation for $H$, and its solution
is
\[ \text{$t = \int \frac{2 d H}{A e^{2 \gamma H} - 3 H^2}$} . \]
Unfortunately, the integral on the right hand side cannot be integrated using
analytic methods.

We now examine the acceleration of the universe in the two cases depending on
the sign of $\gamma$. We can see that $\dot{H} + H^2 = - \frac{1}{2} H^2 +
\frac{A}{2} e^{2 \gamma H}$, so the universe accelerates iff $A e^{2 \gamma H}
> H^2$. In the case $\gamma < 0$, we have $\dot{\rho}_m < 0$ and
$\dot{\rho}_{\Lambda} > 0$. Since the acceleration takes the same sign as $2
\rho_{\Lambda} - \rho_m$, the acceleration occurs as a faster rate as time
proceeds, just as in the non-interacting case. 

In the case $\gamma > 0$,
$\dot{\rho}_{\Lambda} < 0$ always, but the behaviour of $\rho_m$ is more
complicated. A representative example with $\gamma = 1$ is given in Figure 1;
for other values of $\gamma$, the qualitative behaviour is similar (in the
sense that $\rho_m$ generally increases to a maximum and then decreases
again). The straight line in the diagram corresponds to $2 \rho_{\Lambda} =
\rho_m$; in the region of phase space above this line the universe is
accelerating, and in the region below this line it is decelerating. It appears
that in this model the universe can only go through at most a single
decelerating phase.

\bigskip

\tmfloat{h}{small}{figure}{\includegraphics{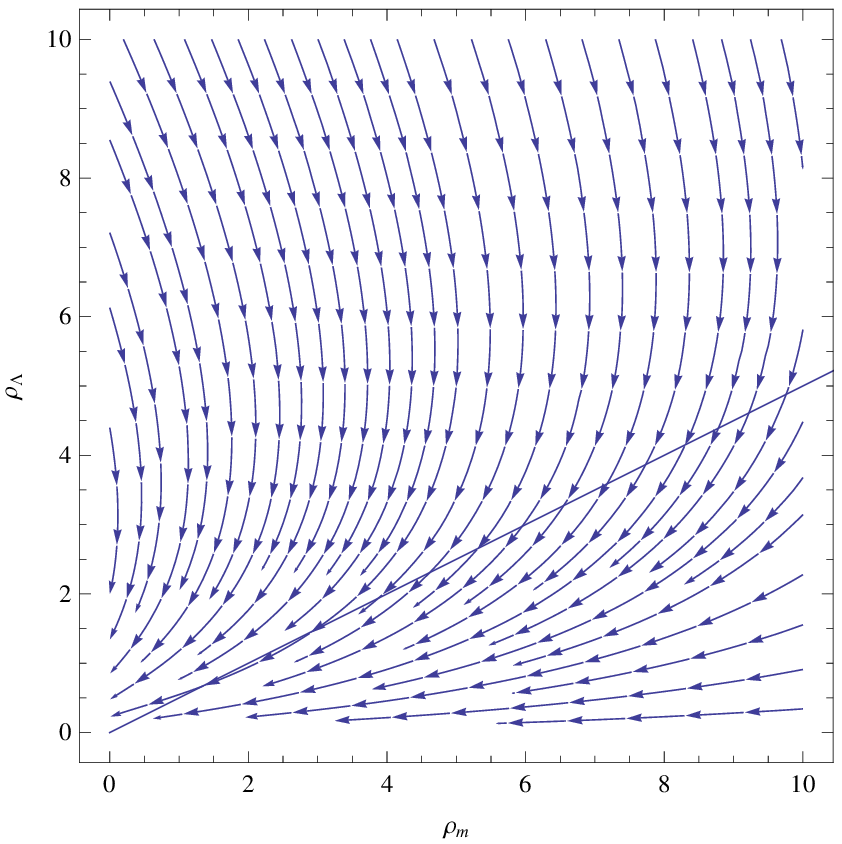}}{Phase-plane diagram showing
the evolution of the dark matter density $\rho_m$ and the dark energy density 
$\rho_{\Lambda}$, for an interaction term $\rho_m\rho_{\Lambda}$ ($\gamma = 1$).
In this case, the dark energy is assumed to behave like a cosmological constant, and obeys
the equation of state $p = -\rho$. The straight line represents universes with zero acceleration. 
The region of phase space above the line corresponds to an accelerating universe, and the region
below corresponds to a decelerating universe. Note that this figure (and the next four) are intended
only to display the qualitative behaviour of the model, so the units on the axes are arbitrary.}

\bigskip

From the expressions above we can see that $r \equiv
\frac{\rho_m}{\rho_{\Lambda}} = \frac{3 H^2}{A e^{2 \gamma H}} - 1$, and this
quantity lies in the interval $[- 1, \frac{3}{A e^2 \gamma^2} - 1]$. (This can
be seen by considering the graph of $y = \frac{3 x^2}{A e^{2 \gamma x}} - 1$.)
Thus, in order for the model to be feasible at all, we require the constants
to satisfy the constraint $A \gamma^2 < 3 e^{- 2}$.

In order for this model to address the coincidence problem, we require that $r
\sim O (1)$, and thus $H^2$ and $e^{2 \gamma H}$ should be of comparable
magnitude. We argue, however, that this cannot happen.

In the case $\gamma < 0$, it is easy to see that the ratio $r$ always
decreases with time, because $\dot{\rho}_m \leqslant 0$ and
$\dot{\rho}_{\Lambda} \geqslant 0$ at all times. Consider, then, the case
$\gamma > 0$. Suppose $\rho_m > c \geqslant 0$ at all times. Then $H$ will
decrease at a rate of at least $\frac{c}{2}$ per unit time (by (\ref{A4})),
and will reach 0 eventually -- but this corresponds to $\rho_m = 0$, which is
a contradiction. Therefore $\rho_m$ must approach 0 asymptotically at late
times, so that the quantity $3 H^2 - A e^{2 \gamma H}$ becomes arbitrarily
small, and comparable to $\rho_{\Lambda} = A e^{2 \gamma H}$ (which must
therefore also be arbitrarily small). Therefore $3 H^2$, and therefore $H$,
becomes arbitrarily small. This is impossible because $3 H^2 - A e^{2 \gamma
H} \rightarrow - A$ as $H \rightarrow 0$.

Hence, this model cannot alleviate the coincidence problem.

\begin{center}
\textbf{V. GENERAL DARK ENERGY}
\end{center}

Now suppose that the dark energy has equation of state $p = w \rho$, where $w$
is a constant, and define $K \equiv w + 1$. The previous case corresponds to
$K = 0$, and the case in Section 3 (two dust fluids) can be regarded as the
case $K = 1$. The general interacting case is then described by the following
equations:
\begin{eqnarray}
  \dot{\rho}_m & = & - 3 H \rho_m + \gamma \rho_m \rho_{\Lambda}  \label{R1}\\
  \dot{\rho}_{\Lambda} & = & - \gamma \rho_m \rho_{\Lambda} - 3 K H
  \rho_{\Lambda}  \label{R2}\\
  3 H^2 & = & \rho_m + \rho_{\Lambda}  \label{R3}
\end{eqnarray}
We assume that $\gamma \neq 0$. Adding (\ref{R1}) and (\ref{R2}) yields
\begin{equation}
  \text{$\dot{\rho}_m + \dot{\rho}_{\Lambda} = - \sqrt{3 (\rho_m +
  \rho_{\Lambda})} (\rho_m + K \rho_{\Lambda})$} \label{R4}
\end{equation}
and subtracting appropriate multiples of (\ref{R1}) and (\ref{R2})
yields
\begin{equation}
  \dot{\rho}_m K \rho_{\Lambda} - \gamma K \rho_m \rho_{\Lambda}^2 =
  \dot{\rho}_{\Lambda} \rho_m + \gamma \rho_m^2 \rho_{\Lambda} .\label{R5}
\end{equation}
Combining (\ref{R4}) and (\ref{R5}), and integrating, gives the
following relation between the dark energy and dark matter densities:
\begin{equation}
  \text{ $\rho_{\Lambda} = \rho_m^K e^{2 \gamma \sqrt{\rho_m + \rho_{\Lambda}}
  / \sqrt{3}}$.}
\end{equation}
Due to this rather awkward relation between the two densities, it is difficult
to make further progress using purely analytical methods. However, we can
still examine the cosmic evolution by treating (\ref{R1}) -- (\ref{R3}) as a
two-dimensional dynamical system:
\begin{eqnarray*}
  \dot{\rho}_m & = & - \sqrt{3} \sqrt{\rho_m + \rho_{\Lambda}} \rho_m + \gamma
  \rho_m \rho_{\Lambda}\\
  \dot{\rho}_{\Lambda} & = & - \gamma \rho_m \rho_{\Lambda} - \sqrt{3}
  \sqrt{\rho_m + \rho_{\Lambda}} K \rho_{\Lambda} .
\end{eqnarray*}

We consider the fixed points of this dynamical system.
The values of $\rho_m$ and $\rho_{\Lambda}$ at these fixed points must satisfy
$\rho_m (\gamma \rho_{\Lambda} - \sqrt{3 (\rho_m + \rho_{\Lambda})}) = 0$ and
$\rho_{\Lambda} (\gamma \rho_m + K \sqrt{3 (\rho_m + \rho_{\Lambda})}) = 0$.
It is clear that the origin $(\rho_m, \rho_{\Lambda}) = (0, 0)$ is always a
fixed point. For fixed points other than the origin, we need the other two
factors to both vanish, and this leads to the consideration of the following
four cases based on the signs of $\gamma$ and $K$:

\begin{center}
\textbf{A. Case 1: $K > 0, \gamma > 0$}
\end{center}

There is a single fixed point $(0, 0)$. We can see that $\dot{\rho}_{\Lambda}
\leqslant 0$ always, so the dark energy density never increases. If
$\rho_{\Lambda}$ is sufficiently great to begin with, the matter density
increases and then decreases. The condition $\dot{\rho}_m = 0$ corresponds to
the parabolic nullcline $\rho_m = \frac{\gamma^2 \rho_{\Lambda}^2}{3} -
\rho_{\Lambda}$.

An illustration of the evolution is shown in Figure 2. In this, as well as the
other cases, the actual values of $\gamma$ and $K$ do not seem to affect the
qualitative behaviour of the orbits, although their signs do.

This scenario is unlikely to represent our present universe. Although for most
trajectories there is an initial dark energy dominated phase which gives way
to a dark-matter dominated phase, the model does not exhibit the late dark
energy dominated phase which we are currently experiencing.

\bigskip

\bigskip

\tmfloat{h}{small}{figure}{\includegraphics{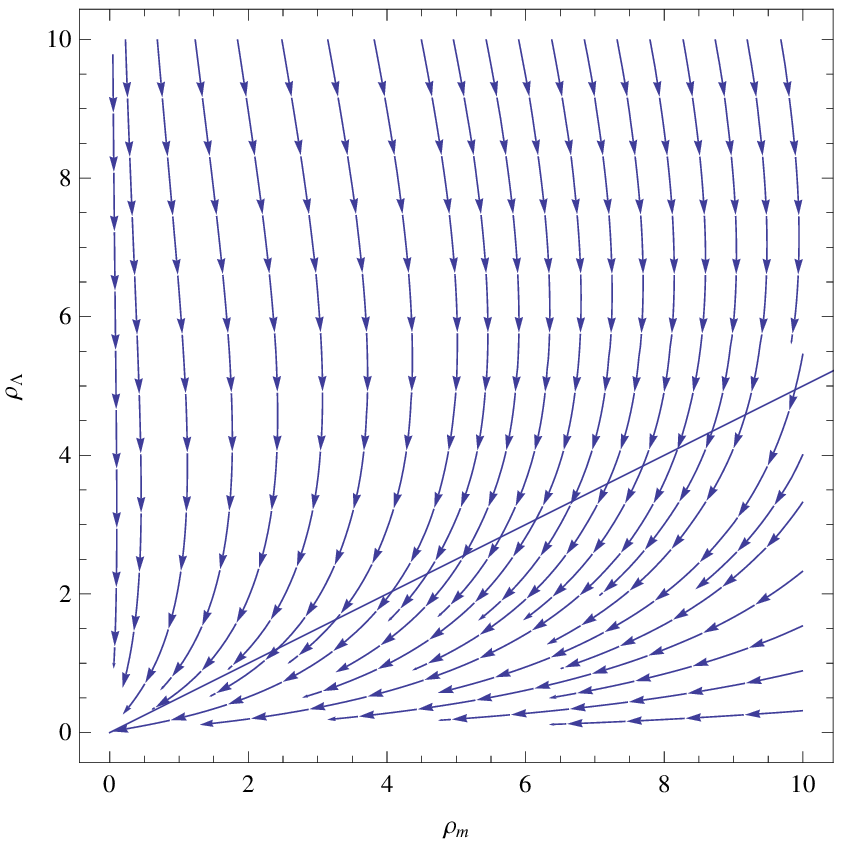}}{As in Figure 1, with the
same interaction term $\rho_m\rho_{\Lambda}$ ($\gamma = 1$), but now assuming a dark energy 
component whose equation of state is $p = -\rho/2$.}

\bigskip

\bigskip

\begin{center}
\textbf{B. Case 2: $K > 0, \gamma < 0$}
\end{center}

There is a single fixed point $(0, 0)$. We can see that $\dot{\rho}_m
\leqslant 0$ always, so the dark matter density never increases. If $\rho_m$
is sufficiently great to begin with, the dark energy density increases and
then decreases. The condition $\dot{\rho}_{\Lambda} = 0$ corresponds to the
parabolic nullcline $\rho_{\Lambda} = \frac{\gamma^2 \rho_m^2}{3 K^2} -
\rho_m$.

In this scenario, an initial matter-dominated phase gives way to a dark energy
dominated phase, and then the density of dark energy falls steeply. However,
this model does not address the coincidence problem, because the ratio $r
\equiv \frac{\rho_m}{\rho_{\Lambda}}$ keeps decreasing (if $K \leqslant 1$). 

\bigskip

\tmfloat{h}{small}{figure}{\includegraphics{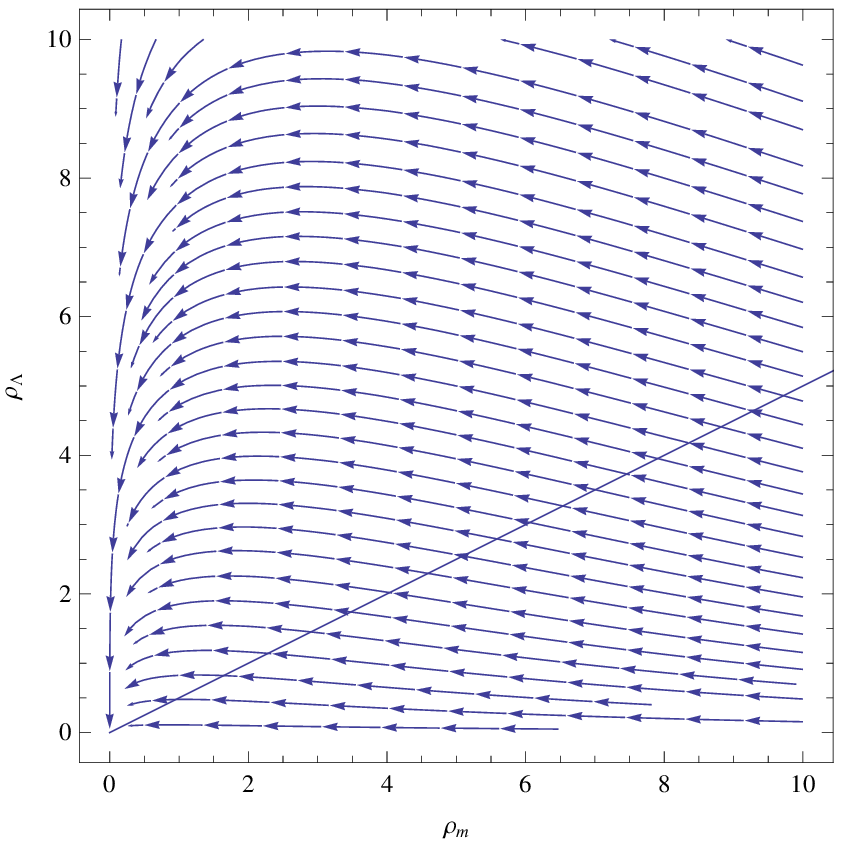}}{As in Figure 1, but with
an interaction term $-\rho_m\rho_{\Lambda}$ ($\gamma = -1$) and a dark energy 
component whose equation of state is $p = -\rho/2$.}

\bigskip

\begin{center}
\textbf{C. Case 3: $K < 0, \gamma < 0$}
\end{center}

In this situation, dark matter transfers energy to `phantom' dark energy. As
usual, there is a single fixed point $(0, 0)$. Since we always have
$\dot{\rho}_m \leqslant 0$ and $\dot{\rho}_{\Lambda} \geqslant 0$, matter is
continually being converted into dark energy, and $\rho_m$ tends monotonically
to 0 (as $\rho_{\Lambda}$ tends monotonically to $\infty$). As the ratio of
the densities $\frac{\rho_m}{\rho_{\Lambda}}$ is always decreasing, this
aggravates the coincidence problem and substantiates the results from [11] and
[12] alluded to earlier in Section 2.

\bigskip

\tmfloat{h}{small}{figure}{\includegraphics{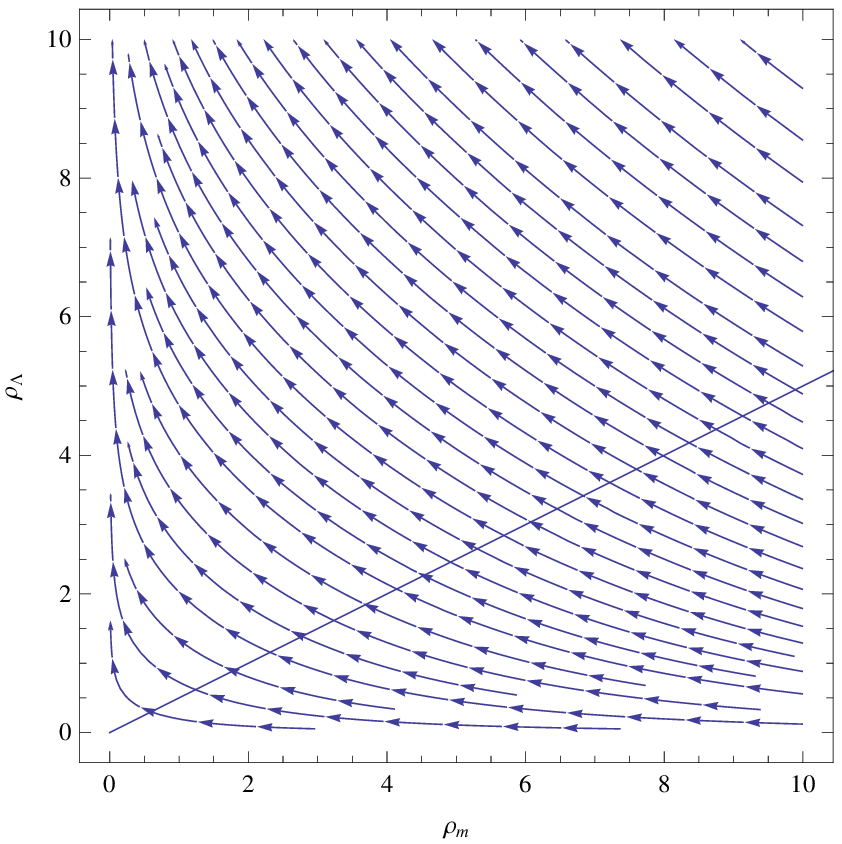}} {As in Figure 1, but with
an interaction term $-\rho_m\rho_{\Lambda}$ ($\gamma = -1$) and a phantom dark energy 
component with equation of state $p = -3\rho/2$.}

\bigskip

\begin{center}
\textbf{D. Case 4: $K < 0, \gamma > 0$}
\end{center}

In this case, `phantom' dark energy transfers energy to dark matter. The
system now has two fixed points at
\[ (0, 0) \quad \mbox{ and } \quad (\bar{\rho}_{m}, \bar{\rho}_{\Lambda}) \equiv \left( \frac{3 K
   (K - 1)}{\gamma^2}, \frac{3 (1 - K)}{\gamma^2} \right) . \]
The eigenvalues of the Jacobian at the second fixed point are $\pm 2 \sqrt{K}
(1 - K)$; they are both purely imaginary, and so $(\bar{\rho}_{m}, \bar{\rho}_{\Lambda})$
is a \tmtextit{center}. Phase-plane plots indicate the existence of
stable periodic orbits around this center, with no convergence to nor
divergence from $(\bar{\rho}_{m}, \bar{\rho}_{\Lambda})$. The
two nullclines are the parabola $\rho_m = \frac{\gamma^2 \rho_{\Lambda}^2}{3}
- \rho_m$ (on which $\dot{\rho}_m = 0$) and the parabola $\rho_{\Lambda} =
\frac{\gamma^2 \rho_m^2}{3 K^2} - \rho_m$ (on which $\dot{\rho}_{\Lambda} =
0$).

A linearization about the centre ($\bar{\rho}_{m}, \bar{\rho}_{\Lambda}$) gives the
equations
\begin{eqnarray*}
  \dot{x} & = & \frac{3 K}{2 \gamma} x + \frac{3 K (2 K - 1)}{2 \gamma} y\\
  \dot{y} & = & \frac{3 (K - 2)}{2 \gamma} x - \frac{3 K}{2 \gamma} y
\end{eqnarray*}
where we have defined $x = \rho_m - \bar{\rho}_{m}$, $y = \rho_{\Lambda} -
\bar{\rho}_{\Lambda}$ and made the assumption that $x, y$ are both small. Fitting
the equation of a conic section $A x^2 + B y^2 + x y = \tmop{const}$ and
differentiating gives a solution $A = \frac{1}{K} - \frac{1}{2}$, $B = K -
\frac{1}{2}$. In order for the trajectories to be ellipses, we require that $4
A B - 1 > 0$, but this is true (as expected) since $4 A B - 1 = 2 \left( \sqrt{- K} +
\frac{1}{\sqrt{- K}} \right)^2$.

Thus, for all $K < 0$, the orbits of the system are bounded. For a given
trajectory, the ratio $\frac{\rho_m}{\rho_{\Lambda}}$ will oscillate between
two extremes given by the gradients of the tangents from the origin to this
trajectory. Note that the interval defined by these two extremes contains the
value $\frac{\bar{\rho}_{\Lambda}}{\bar{\rho}_{m}} = - \frac{1}{K}$, which is $O (1)$
if $K$ is not too close to $0$. In such a model the dark energy and dark
matter will be comparable at infinitely many times, and this would provide a
solution to the coincidence problem. An example of the evolution is shown in
Figure 5 for the parameter values $\gamma = 1$ and $K = - \frac{1}{2}$.

\bigskip

\tmfloat{h}{small}{figure}{\includegraphics{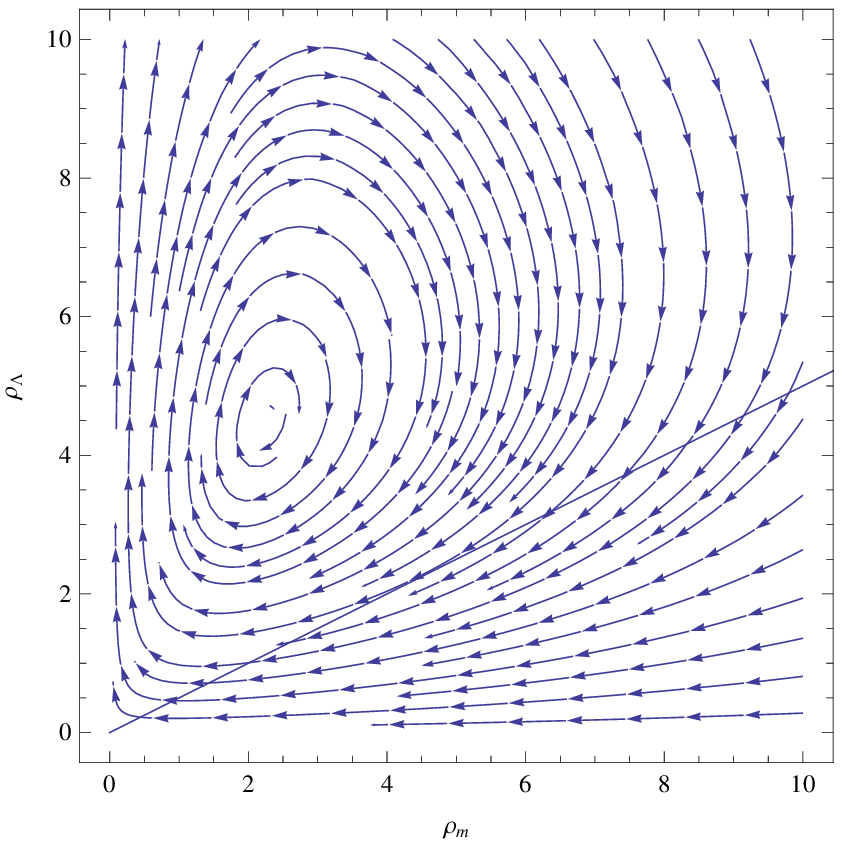}}{As in Figure 1, with
an interaction term $\rho_m\rho_{\Lambda}$ ($\gamma = 1$) but a phantom dark energy 
component with equation of state $p = -3\rho/2$. Note that the system now exhibits 
periodic orbits, and that the ratio of the dark densities for any particular orbit is 
constrained to lie within a bounded interval.}

\begin{center}
\textbf{VI. CONSTRAINING THE MODEL PARAMETERS}
\end{center}

In the previous two sections, we have seen that, in the class of interacting
models with an interaction of the form $\gamma \rho_m \rho_{\Lambda}$, the
only model that could possibly help to alleviate the late-time coincidence
problem is the one in Section V.D. Thus, we will now consider this model in
more detail.

The analysis in the previous sections holds only at late times, when the
baryon and radiation densities are small and can be neglected. In order to
constrain the parameters $\gamma$ and $K$ of the model using observations of
the past universe, we need to examine its past evolution. It is therefore
necessary to include baryons and radiation in the analysis, since their densities 
are significant at early times.

Before we proceed, we should emphasize that the following discussion is 
not an attempt to obtain the best-fit parameters, but merely to show that the 
model under consideration is a viable fit to observations for non-zero $\gamma$ 
and $K$. We have seen that the strength of this model is that it addresses the
coincidence problem (which the standard concordance model cannot do), and we will find
that it appears to fit observations at least as well as the concordance model. In 
order to obtain best-fit parameters and make a more detailed comparison to 
observations, more careful work is needed, including a full stability analysis 
of perturbations. This is, however, outside the scope of the present paper, and is
left as a topic for future investigation.

The complete system of equations to be solved is:
\begin{eqnarray*}
  (1 + z) \rho'_m & = & 3 \rho_m - \frac{\gamma \rho_m \rho_{\Lambda}}{H} \label{Z1}\\
  (1 + z) \rho'_{\Lambda} & = & 3 K \rho_{\Lambda} + \frac{\gamma \rho_m  \rho_{\Lambda}} {H}  \label{Z2}\\
  3 H^2 & = & 8 \pi G (\rho_m + \rho_{\Lambda} + \rho_{b 0} (1 + z)^3 +
  \rho_{r 0} (1 + z)^4) \label{Z3}
\end{eqnarray*}
where we have transformed the independent coordinate from cosmic time $t$ to
redshift $z$, and a prime indicates a derivative with respect to $z$. The
quantities $\rho_{b 0}$ and $\rho_{r 0}$ are the present densities of baryons
and photons, respectively.

In order to find suitable choices of parameters for $\gamma$ and $K$, we can
integrate these equations numerically for various values of these parameters,
and compare the results to observations. In particular, we shall attempt to
examine the effect of three observational constraints which are claimed to be
model-independent [72]: (i) the shift parameter $R$, related to the angular scale
of the first acoustic peak in the CMB power spectrum, (ii) the distance
parameter $A$, related to the measurement of the BAO peak from a sample of
SDSS luminous red galaxies, and (iii) Type IA Supernovae data.

The shift parameter $R$ is defined as follows [72, 73]:
\[ R \equiv (\Omega_{m 0} + \Omega_{b 0})^{1 / 2} H_0
   \int_0^{z_{\tmop{recomb}}} \frac{d z'}{H (z')} \]
where $z_{\tmop{recomb}} = 1091.3$ [74] (WMAP7) is the recombination redshift. When testing
our model, we will adopt the values $\Omega_{m 0} = 0.227$ for the present fractional density of dark matter, 
$\Omega_{b 0} = 0.0456$ for the present fractional density of baryons, and $H_0 = 70.4 km$ s$^{- 1}$ Mpc$^{- 1}$ 
for the present value of the Hubble parameter. Note that this is only an approximation, because the 
derivation of these values from the WMAP7 data assumes a standard, non-interacting $\Lambda$CDM model.
The value of $R$ obtained from the WMAP7 data is $1.725 \pm 0.018$. 

The distance parameter $A$ is defined as follows [75, 76]:
\[ A \equiv (\Omega_{m 0} + \Omega_{b 0})^{1 / 2} \frac{H_0}{H (z_b)^{1 / 3}} \left[
   \frac{1}{z_b} \int_0^{z_b} \frac{d z'}{H (z')} \right]^{2 / 3} \]
where $z_b = 0.35$. The value of $A$ has been determined to be $0.469 (n_s /
0.98)^{- 0.35} \pm 0.017$ ($1 \sigma$ constraints) [76], and we will take the
scalar spectral index $n_s$ to be $0.963$ [74] (WMAP7). We will also assume the WMAP7 parameters
for $\Omega_{m 0}$ and $\Omega_{b 0}$ (discussed above). 

Using these parameters, we have calculated $R$ and $A$ for various values of
$\gamma > 0$ and $K < 0$. The observational constraints from $R$ put a crude upper bound on
$\gamma$ of about $1.4 \times 10^7$ m$^3$ kg$^{- 1}$ s$^{- 1}$, and requires
$K > -0.48$. The constraint from $A$ suggests that $K > - 0.27$ and 
requires $\gamma$ to be less than about $1.8 \times 10^9$ m$^3$ kg$^{- 1}$ s$^{- 1}$. The closer $K$ is 
to 0, the higher the values of $\gamma$ permitted by either test: a more negative
value of $K$ tends to increase the values of both $A$ and $R$.

In the remainder of this paper, we will take the parameters $\gamma = 100$
and $K = - 0.15$. Note that, although these are not necessarily the best-fit
parameters, they are in accord with the observed values of $R$ and $A$.

We have fitted this particular model to SNIa data [77]. At
the outset, we might expect that there will not be a large deviation from the
behaviour of the concordance model in this regime, since the densities of
either dark matter or dark energy would be very low at small redshifts, and
the interaction term would therefore be negligible. The recent evolution of the system
would therefore be similar to that of a standard $\Lambda$CDM universe. The results confirm this: for a
sample of 608 Union2 supernovae, the model with $\gamma = 100, K = -0.15$ gives a
$\chi^2$ value of $756.4$, which is slightly better than the one obtained for the
$\gamma = 0, K = 0$ model ($780.7$).

\begin{center}
\textbf{VII. DISCUSSION}
\end{center}

We now consider the concrete instance of our model with $\gamma = 100$ and $K
= - 0.15$. The evolution of the fractional densities with redshift is shown in Figure 6:

\bigskip

\tmfloat{h}{small}{figure}{\resizebox{12cm}{8cm}{\includegraphics{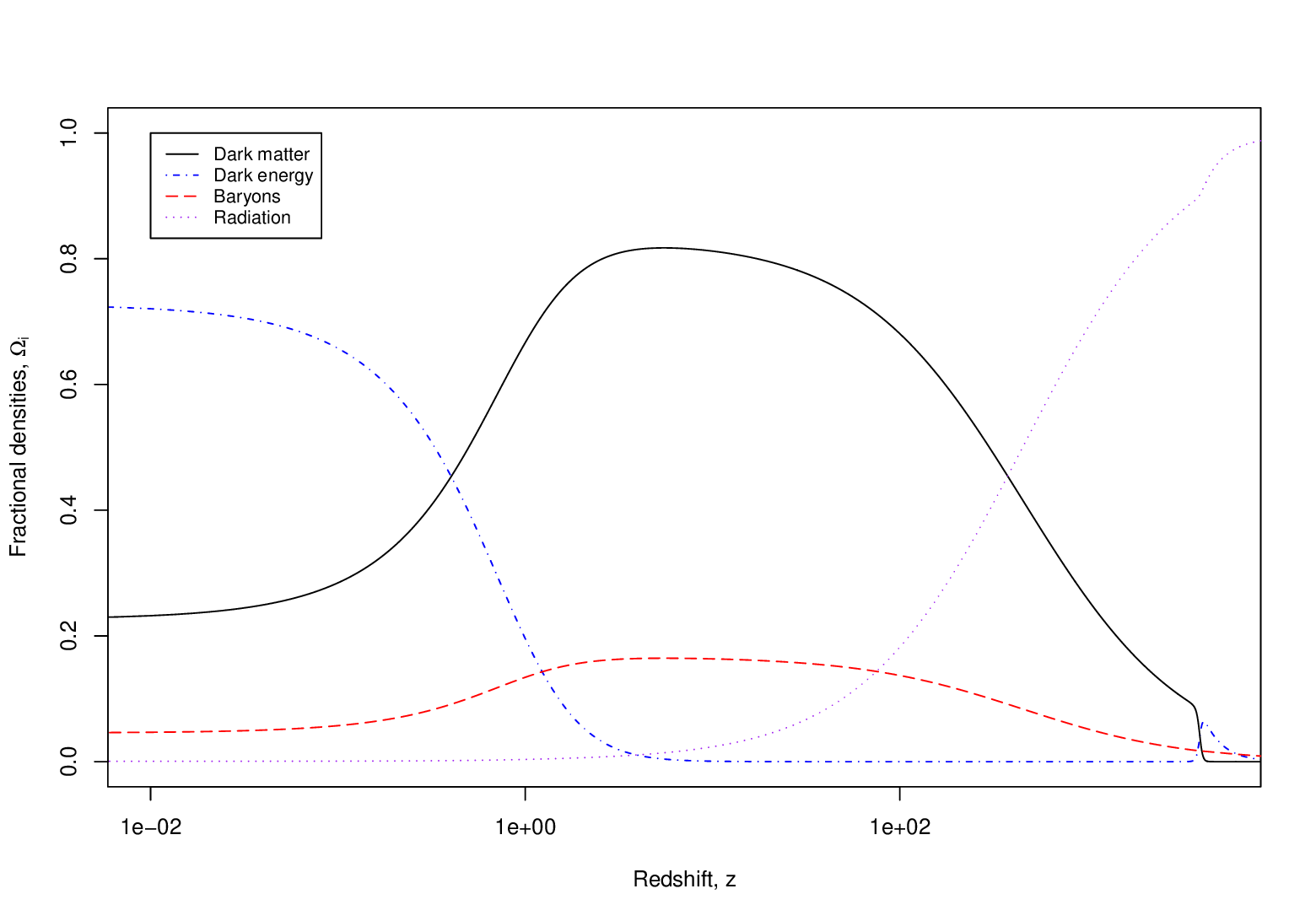}}}{The past
evolution of the fractional densities $\Omega_i$ with redshift, for a universe in which $\gamma =
100$ and $K = - 0.15$. The universe undergoes the usual sequence of radiation, dark matter and dark energy
dominated eras.}

\bigskip

From Figure 6, we can see that this model provides a suitable sequence of
cosmological eras corresponding to what we know about our universe. The
universe is radiation-dominated at very early times. This is followed by a long
matter-dominated era, with $\Omega_{\Lambda} \ll \Omega_m$, up to around
redshift $z \approx 0.5$. At late times, we encounter a period of
dark-energy-dominated acceleration.

The past history of the universe is somewhat dependent on the parameter $\gamma$. In 
particular, increasing $\gamma$ to larger than about $10^4$ induces a peak in the baryon fractional
density at about redshift 2500 and leads to a brief baryon-dominated era around that time,
whilst moving the transition from the radiation-dominated era to the matter-dominated era later in time (i.e.,
to smaller redshift).

Also, note the peak in the fractional density of dark energy at around 
$z \approx 2500$. This corresponds to the beginning of a periodic cycle. There are, in
fact, infinitely many such cycles, as the evolution continues into the past, and the oscillations
get more rapid as we get closer and closer to the Big Bang (since, at early
times, the Hubble parameter is large due to the high density of radiation).
However, this is not easy to see even if we extrapolate Figure 7 to higher
redshifts, since at high redshifts $\rho_m$ and $\rho_{\Lambda}$ are swamped
by the radiation density. We can, however, observe this effect on a
logarithmic plot (Figure 7):

\bigskip

\tmfloat{h}{small}{figure}{\resizebox{12cm}{8cm}{\includegraphics{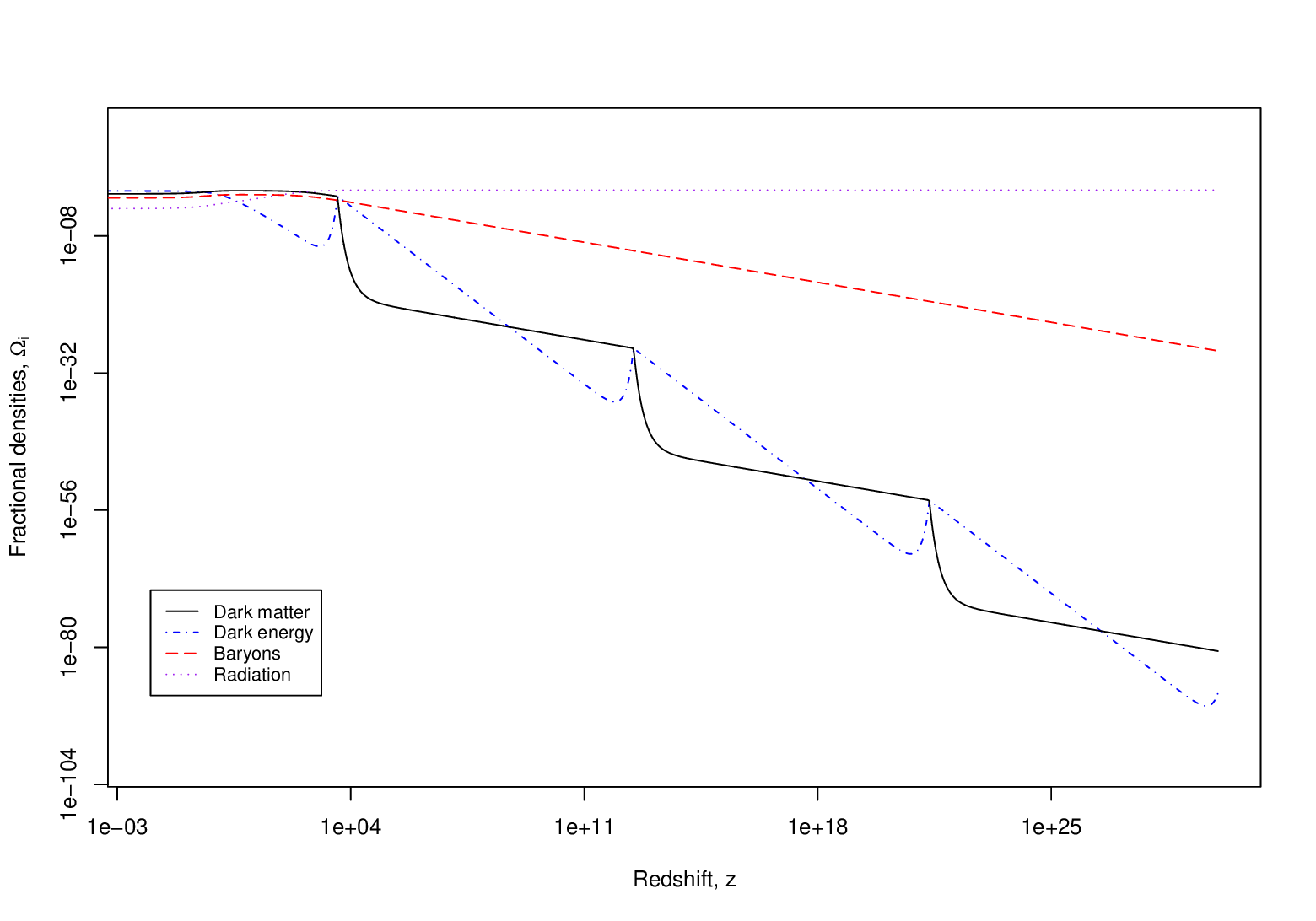}}}{A
log-log plot of the fractional densities $\Omega_i$ against redshift, for a universe with $\gamma = 100$
and $K = -0.15$. This plot shows that many cycles may occur before the current one.}

\bigskip

Having established that the model adequately describes the past universe, let
us consider its implications for the late universe. At late times, the
densities of the non-interacting baryons and photons will decrease and become
negligible. The evolution of the universe thus converges to a stable limit
cycle, in which the following four phases repeatedly occur:
\begin{itemizedot}
  \item Dark energy is converted quickly into dark matter, while $H$ remains large.
  
  \item When the density of dark energy is sufficiently low, the interaction
  term in the evolution equation for $\dot{\rho}_m$ is overwhelmed by the $- 3
  H \rho_m$ term, and the density of dark matter starts to decrease, while the
  dark energy density stays low. The value of $H$ decreases, and the universe decelerates. 
  
  \item The dark matter and dark energy densities are small and comparable. The dark
  energy density slowly starts to increase while the dark matter density
  continues to decrease. This marks the transition from a dark-matter-dominated universe 
  to a dark-energy-dominated universe, and the universe begins to accelerate. The universe 
  that we are currently experiencing is in the later stages of this transition.
  
  \item When the density of dark energy then becomes sufficiently large, it continues to increase
  until the interaction term becomes non-negligible. 
\end{itemizedot}
This process goes on forever, with eternally alternating dark-matter- and
dark-energy-dominated eras. Also, the universe undergoes infinitely many periods
of acceleration and deceleration, as can be seen from equation (\ref{ACCEL}).

In this limit cycle, the dark matter and dark energy densities are comparable
for a significant proportion of the time. The following graph shows a plot demonstrating this:

\bigskip

\tmfloat{h}{small}{figure}{\resizebox{12cm}{8cm}{\includegraphics{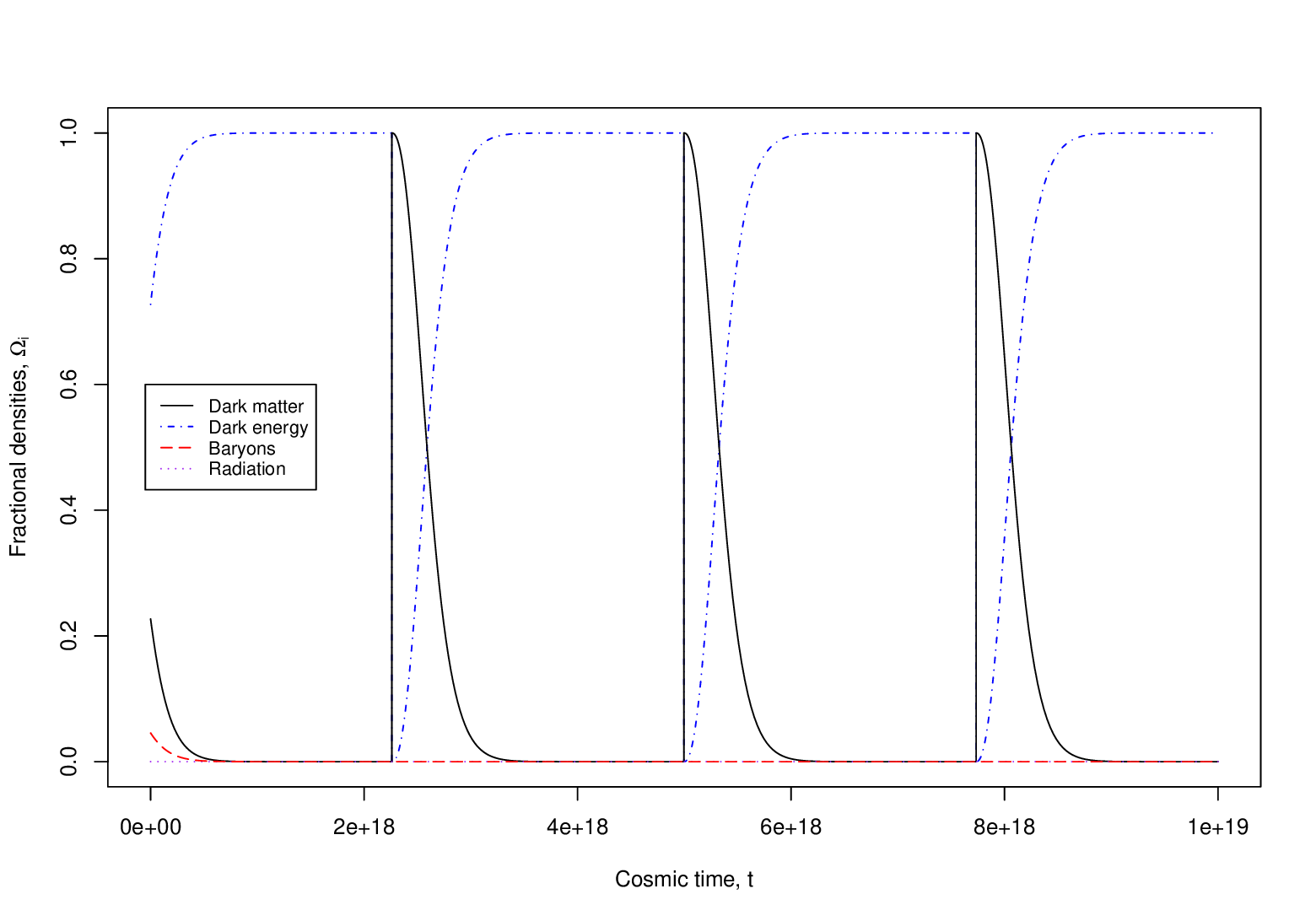}}}{The
future evolution of the fractional densities $\Omega_i$ with cosmic time $t$, for $\gamma =
100$ and $K = - 0.15$. In this plot, $t$ is measured in seconds, and the time $t = 0$ corresponds to the present.}

\bigskip

The duration of an entire cycle is dependent on the choice of parameters:
choosing a more negative value of $K$ leads to more and shorter cycles. From
Figure 8, the densities are comparable for about a quarter of each cycle, and
that dark energy is dominant at most other times. This suggests that, in this
model, it is not unnatural for the universe to be in a state at which the dark
matter and dark energies are comparable.

This model, however, has a limitation: it does not explain the apparent coincidence of the
baryon and dark matter densities being comparable at the present time. In particular, the baryon density
continues to decrease steadily as the cycles progress, whereas the minimum value of the dark matter 
density stays at roughly the same level over all future cycles, because of its periodic interaction with 
dark energy.

\begin{center}
\textbf{VIII. CONCLUSIONS}
\end{center}

We have considered a model with an interaction term proportional to the
product of the densities of the interacting fluids, and applied it to the
interaction of a dust fluid and another fluid with equation of state $p = w
\rho$, for various values of the constant $w$. In the case of two dust fluids,
we have obtained a general solution, and have shown that their density
evolution is similar to that in the non-interacting case, after an initial
short cooling-off period.

We then proceeded to show that such an interaction between dark matter and
dark energy in the form of a cosmological constant would not solve the
coincidence problem. In the case $\gamma < 0$, energy is transferred from dark
matter to dark energy, and the ratio of the energy densities $r = \rho_m /
\rho_{\Lambda}$ always decreases with time. We would therefore expect the dark
matter density today to be negligible compared to the dark energy density.
However, this does not correspond to what we observe in the universe today. On the
other hand, in the case $\gamma > 0$ where energy is transferred from dark
energy to dark matter, we found that it was still not possible to have the
densities be of comparable magnitude. This led to the conclusion that such an
interaction would not help to solve the coincidence problem. The same is true
if $\gamma < 0$ and $w$ is allowed to vary arbitrarily, or if $w > - 1$ and
$\gamma$ is allowed to vary arbitrarily.

However, it is interesting to note that if the dark energy follows a `phantom'
equation of state ($w < - 1$) and energy is transferred from dark energy to
dark matter, it is possible to obtain periodic orbit solutions. These
correspond to cyclic situations in which the ratio of the dark densities is
comparable at infinitely many times. We have also shown that suitable parameters
$\gamma$ and $K$ can be chosen so that the model is consistent with observations.

Such a scenario could alleviate the coincidence problem, because it can be
shown that if $w$ is not too close to $- 1$, the periodic orbits would enclose
a fixed point corresponding to a density ratio $r$ that is $O (1)$, and thus
the value of $r$ on these trajectories would be $O (1)$ at infinitely many
times in the evolution of the universe.

\begin{center}
\textbf{ACKNOWLEDGMENTS}
\end{center}

I would like to thank my supervisor, John Barrow, for helpful comments; Simeon Bird, 
David Essex, Hiro Funakoshi, Baojiu Li, Yin-Zhe Ma and an anonymous referee
for useful discussions; and the Gates Cambridge Trust for its support.

\end{document}